# A new measurement method of electrode gains for orthogonal symmetric type beam position monitor*

ZOU Jun-Ying(邹俊颖)[1)] WU Fang-Fang(吴芳芳) YANG Yong-Liang(杨永良) SUN Bao-Gen(孙葆根)[2)]
Zhou Ze-Ran(周泽然) Luo Qing(罗箐) LU Ping(卢平) XU Hong-Liang(徐宏亮)

NSRL, School of Nuclear Science and Technology,
University of Science and Technology of China, Hefei 230029, P. R. China

**Abstract:** The new beam position monitor (BPM) system of the injector at the upgrade project of Hefei Light Source (HLS II) has 19 stripline beam position monitors. Most consist of four orthogonal symmetric stripline electrodes. The differences in electronic gain and mismachining tolerance can cause the change of the beam response of the BPM electrodes. This variation will couple the two measured horizontal positions in order to bring the measuring error. To alleviate this effect, a new technique to measure the relative response of the four electrodes has been developed. It is irrelevant to the beam charge and the related coefficient can be theoretical calculated. The effect of electrodes coupling on this technique is analyzed. The calibration data is used to fit the gain for all 19 injector beam position monitors. The results show the standard deviation of the distribution of measured gains is about 5%.

**Key words:** beam position monitor, electrode gain, calibration, orthogonal symmetric
**PACS:** 29.20.Ej, 29.90.+r

## 1 Introduction

Recently, Hefei Light Source (HLS) is being upgraded to HLS II. The injector beam position monitoring (BPM) system is composed of 19 beam position monitors, mostly are regular stripline type BPM. They are precisely calibrated and carefully installed in place [1]. We have developed a new technique that provides a measure of the relative gain of the four stripline electrodes.

The method we developed is similar to the technique of D.L. Rubin [2] et al. It also based on the fact that, in a four electrodes beam position monitor, the position of the beam is overdetermined. The relative gains of the electrodes can be calculated by measuring the electrode signal at many different beam positions. The method of Rubin is based on the image theory, which requires the geometry of the four BPM electrodes be diagonal symmetric. The geometry of a typical HLS II beam position monitor is as in Fig. 1. The four electrodes are orthogonal symmetric, which does not apply to Rubin's method, so we develop a new technique to measure the relative gains of this type of four electrodes beam position monitor. Through the analysis of the theoretical electrode signal induced by the beam, we find a new expression only related to the electrode signal. This expression can be used to fit the electrode gain errors, within each fitting procedure, four unknown parameters are fitted: three button gains and a geometry scaling factor.

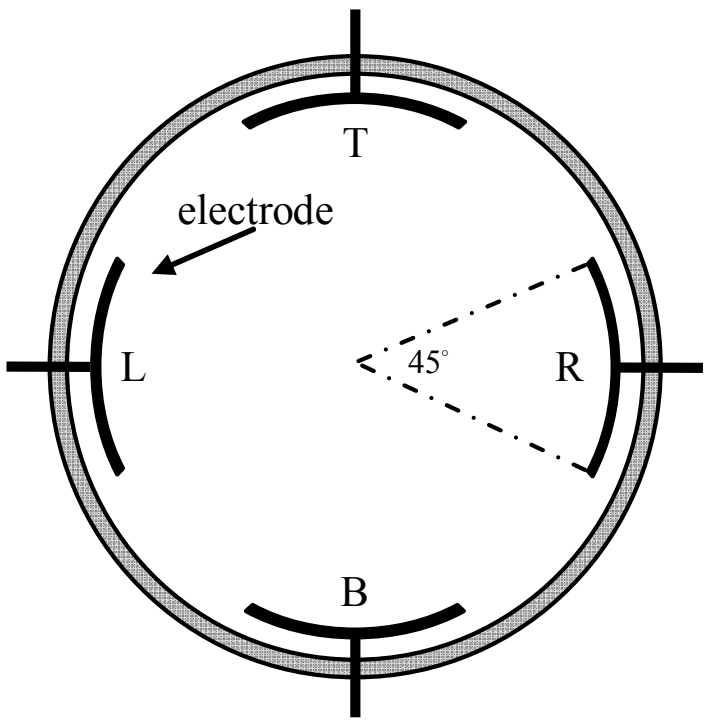

Fig. 1. HLS II injector beam position monitor

* Supported by the Natural Science Foundation of China (11175173, 11375178, 11105141)

1) E-mail: zoujyyl@mail.ustc.edu.cn

2) E-mail: bgsun@ustc.edu.cn

## 2 Derivation of new expression

As Fig. 1 shown, the four electrodes of a HLS II typical BPM are 90 degrees away from each other. By ignoring the influence of bunch size, the electrode signal of this type of BPM can be represented by [3]

$$\begin{cases} V_R = \frac{I_{beam}\phi}{2\pi b}\left(1+Z_{1x}+Z_2+Z_{3x}+Z_4+...\right) \\ V_L = \frac{I_{beam}\phi}{2\pi b}\left(1-Z_{1x}+Z_2-Z_{3x}+Z_4+...\right) \\ V_T = \frac{I_{beam}\phi}{2\pi b}\left(1+Z_{1y}-Z_2-Z_{3y}+Z_4+...\right) \\ V_B = \frac{I_{beam}\phi}{2\pi b}\left(1-Z_{1y}-Z_2+Z_{3y}+Z_4+...\right) \end{cases} \quad (1)$$

Which, $I_{beam}$ is the beam charge, $\phi$ is the electrodes opening angle, $b$ is the distance from center of the beam position monitor to the electrodes. $Z_{1x}$, $Z_{1y}$, $Z_2$, $Z_{3x}$, $Z_{3y}$ and $Z_4$ are introduced in order to simplify the expressions

$$\begin{cases} Z_{1x} = 2\frac{\sin(\phi/2)}{\phi/2}\frac{x_0}{b}, Z_{1y} = 2\frac{\sin(\phi/2)}{\phi/2}\frac{y_0}{b}, \\ Z_2 = 2\frac{\sin\phi}{\phi}\frac{x_0{}^2-y_0{}^2}{b^2}, \\ Z_{3x} = 2\frac{\sin(3\phi/2)}{3\phi/2}\frac{x_0{}^2-3y_0{}^2}{b^2}\frac{x_0}{b}, \\ Z_{3y} = 2\frac{\sin(3\phi/2)}{3\phi/2}\frac{3x_0{}^2-y_0{}^2}{b^2}\frac{y_0}{b}, \\ Z_4 = \frac{\sin(2\phi)}{\phi}\frac{3\left(x_0{}^2-y_0{}^2\right)^2-2\left(x_0{}^4+y_0{}^4\right)}{b^4}. \end{cases} \quad (2)$$

Which, $x_0$ and $y_0$ are the positions of the beam. When the beam is near the center of the beam pipe, $x_0$ and $y_0$ are small compared to $b$. In this case, the third order and up can be ignored, so the electrode signals can be approximated as a quadratic polynomial expansion

$$\begin{cases} V_R = \frac{I_{beam}\phi}{2\pi b}(1+Z_{1x}+Z_2) \\ V_T = \frac{I_{beam}\phi}{2\pi b}(1+Z_{1y}-Z_2) \\ V_L = \frac{I_{beam}\phi}{2\pi b}(1-Z_{1x}+Z_2) \\ V_B = \frac{I_{beam}\phi}{2\pi b}(1-Z_{1y}-Z_2) \end{cases} \quad (3)$$

Taking sums and differences of Eq. (3) gives

$$\begin{cases} \Sigma_{mn} = \frac{V_R+V_L-\left(V_T+V_B\right)}{V_R+V_L+V_T+V_B} = Z_2 \\ \Sigma_m = \frac{V_R-V_L}{V_R+V_L} - \frac{V_T-V_B}{V_T+V_B} \\ \quad = \frac{Z_{1x}-Z_{1y}-Z_2Z_{1x}-Z_2Z_{1y}}{1-Z_2Z_2} \\ \Sigma_n = \frac{V_R-V_L}{V_R+V_L} + \frac{V_T-V_B}{V_T+V_B} \\ \quad = \frac{Z_{1x}+Z_{1y}-Z_2Z_{1x}+Z_2Z_{1y}}{1-Z_2Z_2} \end{cases} \quad (4)$$

Also, ignore the third order and up we can simply get

$$\begin{cases} \Sigma_{mn} = Z_2 \\ \Sigma_m = Z_{1x} - Z_{1y} \\ \Sigma_n = Z_{1x} + Z_{1y} \end{cases} \quad (5)$$

Combining Eq. (2) and Eq. (5) to eliminate $x_0$ and $y_0$ gives an expression that simply relates the electrode signals

$$\begin{cases} \Sigma_{mn} = k_{mm}\Sigma_m\Sigma_n \\ k_{mm} = \frac{\phi}{4\tan\left(\phi/2\right)} \end{cases} \quad (6)$$

In this case, $k_{mn}$ is a constant only determined by the electrodes opening angle of BPM. To the regular injector stripline BPM of HLS II, $\phi$ is 45 degree, so we can simply calculate that $k_{mn}$ is 0.474. Eq. (6) not only shows that $\Sigma_{mn}$ is proportional to the product $\Sigma_m\Sigma_n$, more importantly, the equation is irrelevant to the beam charge, which is useful when fit the gain errors using real beam.

## 3 Simulation

To simulate the connection between $\Sigma_{mn}$ and $\Sigma_m\Sigma_n$, we used a finite element code to create a map of each electrode response as a function of beam position [4].

The simulated beam was moved in a 5 *mm* × 5 *mm* square area with a step of 0.5 *mm*. $\Sigma_{mn}$ and $\Sigma_m\Sigma_n$ was calculated with the exact response of electrodes at every beam positions. The product $\Sigma_m\Sigma_n$ is plotted versus $\Sigma_{mn}$ in Fig. 2. In Fig.2, the points deviation from the straight line only slightly appears at large amplitudes, shows the extent to which the higher than second order terms can be ignored.

We see that our quadratic term approximation is good, the product $\Sigma_m\Sigma_n$ approximated to $\Sigma_{mn}$, which fits the form of Eq. (5) with slightly deviation at large amplitudes.

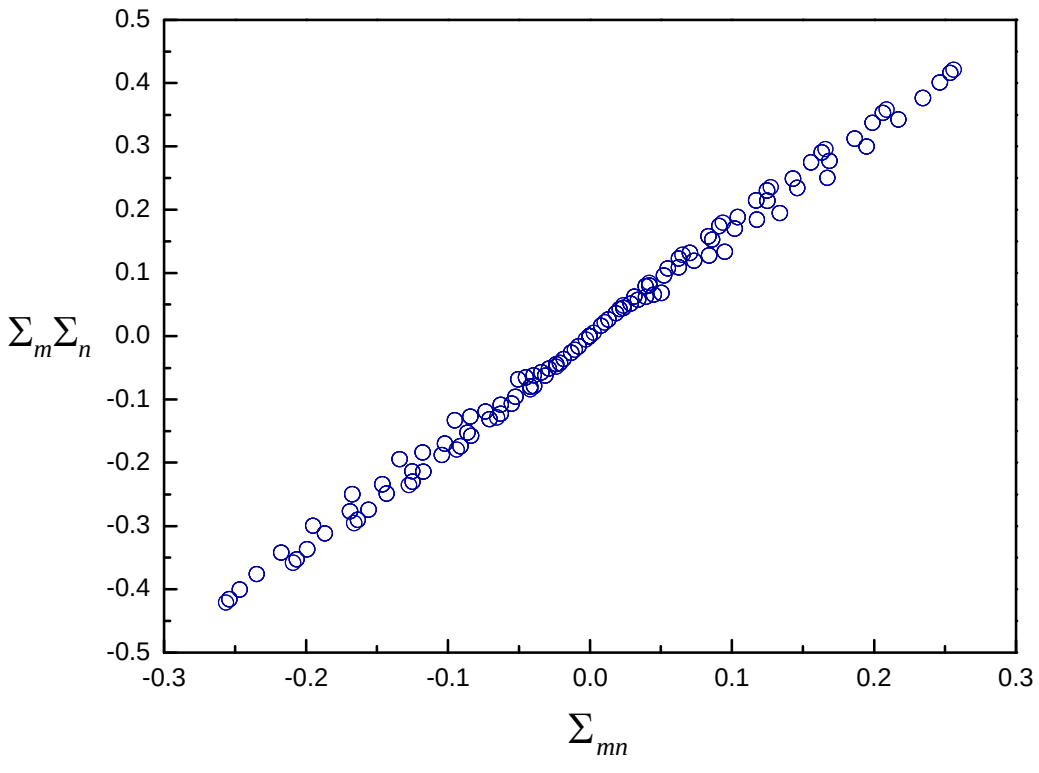

Fig. 2. $\Sigma_m\Sigma_n$ vs $\Sigma_{mn}$ for points on a 5 *mm* ×5 *mm* grid with simulated electrodes signal vs beam position.

In practice, the four electrodes do not have the same gain, then the connection between electrodes defined by Eq. (6) will fail. We simulate the effect of gain errors by reducing the signal on electrodes 4 by 10%, that is, the gains (1:4) = 1.0, 1.0, 1.0, 0.9. Fig. 3 shows the $\Sigma_m\Sigma_n$ vs $\Sigma_{mn}$ with the data under this condition, + indicates the coordinate (0,0). The data is no longer linear and it is offset from zero.

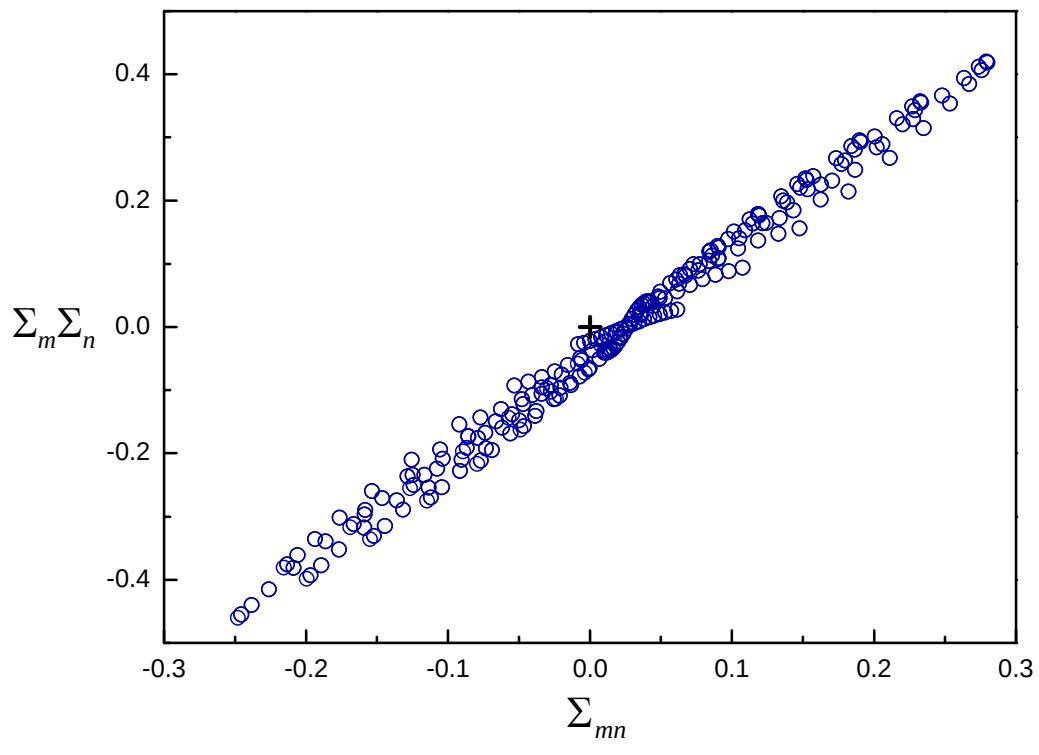

Fig. 3. $\Sigma_m\Sigma_n$ vs $\Sigma_{mn}$ for points on a 5 *mm* ×5 *mm* grid with electrode intensity computed with the nonlinear map.

## 4 Electrodes coupling effect

Eq. (6) is based on the assumption that the four electrodes are independent to each other. In fact, there is coupling effect between the electrodes. Each electrode can be induced to signals from other electrodes. We set $K_1$ as the coupling coefficient of opposite electrode, $K_2$ as the coupling coefficient of adjacent electrode. So the four electrodes signals are given by

$$\begin{cases} \tilde{V}_R = V_R + K_1V_L + K_2V_T + K_2V_B \\ \tilde{V}_L = K_1V_R + V_L + K_2V_T + K_2V_B \\ \tilde{V}_T = K_2V_R + K_2V_L + V_T + K_1V_B \\ \tilde{V}_T = K_2V_R + K_2V_L + K_1V_T + V_B \end{cases} \tag{7}$$

In this case, we calculate Eq. (4) by ignoring the third order and up

$$\begin{cases} \tilde{\Sigma}_{mn} = \dfrac{\tilde{V}_R + \tilde{V}_L - \left(\tilde{V}_T + \tilde{V}_B\right)}{\tilde{V}_R + \tilde{V}_L + \tilde{V}_T + \tilde{V}_B} = \dfrac{1 - 2K_2 + K_1}{1 + 2K_2 + K_1} Z_2 \\ \tilde{\Sigma}_m = \dfrac{\tilde{V}_R - \tilde{V}_L}{\tilde{V}_R + \tilde{V}_L} - \dfrac{\tilde{V}_T - \tilde{V}_B}{\tilde{V}_T + \tilde{V}_B} = \dfrac{(1-K_1)\left(Z_{1x} - Z_{1y}\right)}{(1 + 2K_2 + K_1)} \\ \tilde{\Sigma}_n = \dfrac{\tilde{V}_R - \tilde{V}_L}{\tilde{V}_R + \tilde{V}_L} + \dfrac{\tilde{V}_T - \tilde{V}_B}{\tilde{V}_T + \tilde{V}_B} = \dfrac{(1-K_1)\left(Z_{1x} + Z_{1y}\right)}{(1 + 2K_2 + K_1)} \end{cases} \tag{8}$$

So the Eq. (6) can be modified to

$$\begin{cases} \tilde{\Sigma}_{mn} = \tilde{k}_{mn}\tilde{\Sigma}_m\tilde{\Sigma}_n \\ \tilde{k}_{mn} \approx \dfrac{(1+K_1)^2 - 4K_2^2}{(1-K_1)^2} \dfrac{\phi}{4\tan(\phi/2)} \end{cases} \tag{9}$$

$\tilde{k}_{mn}$ is a coefficient determined by the electrodes coupling effect and the electrodes opening angle. We calculate the coupling coefficients through the analysis of the simulation BPM model using CST-Microwave Studio software. A simulated Gaussian signal is generated at one electrode. By integrating the original signal and the induced signal at other electrodes, we can get $K_1$ is 1.82%, $K_2$ is 5.52%. Finally we get $\tilde{k}_{mn}$ is about 0.504.

## 5 Electrode gain fit with new expression

We assume the deviations from Eq. (9) are determined by the gain variations between different electrodes. We use a nonlinear least squares fit to get the electrode gains ($g_R$, $g_L$, $g_T$ and $g_B$). The merit function to be minimized is

$$\chi^2 = \sum_{i=1}^{n} \left( \begin{aligned} & \frac{g_R\tilde{V}_R + g_L\tilde{V}_L - \left(g_T\tilde{V}_T + g_B\tilde{V}_B\right)}{g_R\tilde{V}_R + g_L\tilde{V}_L + g_T\tilde{V}_T + g_B\tilde{V}_B} \\ & -\tilde{k}_{mn}\left(\frac{g_R\tilde{V}_R - g_L\tilde{V}_L}{g_R\tilde{V}_R + g_L\tilde{V}_L} - \frac{g_T\tilde{V}_T - g_B\tilde{V}_B}{g_T\tilde{V}_T + g_B\tilde{V}_B}\right) \\ & \times\left(\frac{g_R\tilde{V}_R - g_L\tilde{V}_L}{g_R\tilde{V}_R + g_L\tilde{V}_L} + \frac{g_T\tilde{V}_T - g_B\tilde{V}_B}{g_T\tilde{V}_T + g_B\tilde{V}_B}\right) \end{aligned} \right)^2 \tag{10}$$

$\chi^2$ has a minimum for the best fit gains ($g_R$, $g_L$, $g_T$ and $g_B$) and $\tilde{k}_{mn}$. To make sure the value of the denominator is not zero, we fit the same data four times, each time we set one of the electrode gains to 1, and then average the results.

## 6 Fitting the calibration data

All the 19 HLS II injector stripline BPMs are calibrated at test bench, using a tungsten filament to simulate the beam [1]. The filament was moved in a 5 *mm* ×5 *mm* square area with a step of 0.5 *mm*. We collect the electrodes signal data on each simulated beam position using Libera Brilliance Single Pass [5]. An example of fitted data based on Eq. (10) at one BPM (LA-BD-BPM03) is shown in Fig. 4. In Fig. 4, the open circles are the raw electrode data, the crosses are the electrode data corrected with the fitted gains, the + indicates the coordinate (0,0). The fitted gains($g_R$, $g_L$, $g_T$ and $g_B$) respectively are 0.882, 1.122, 0.923 and 1.122. The result shows the data has better linearity and passes through zero after gain fitting.

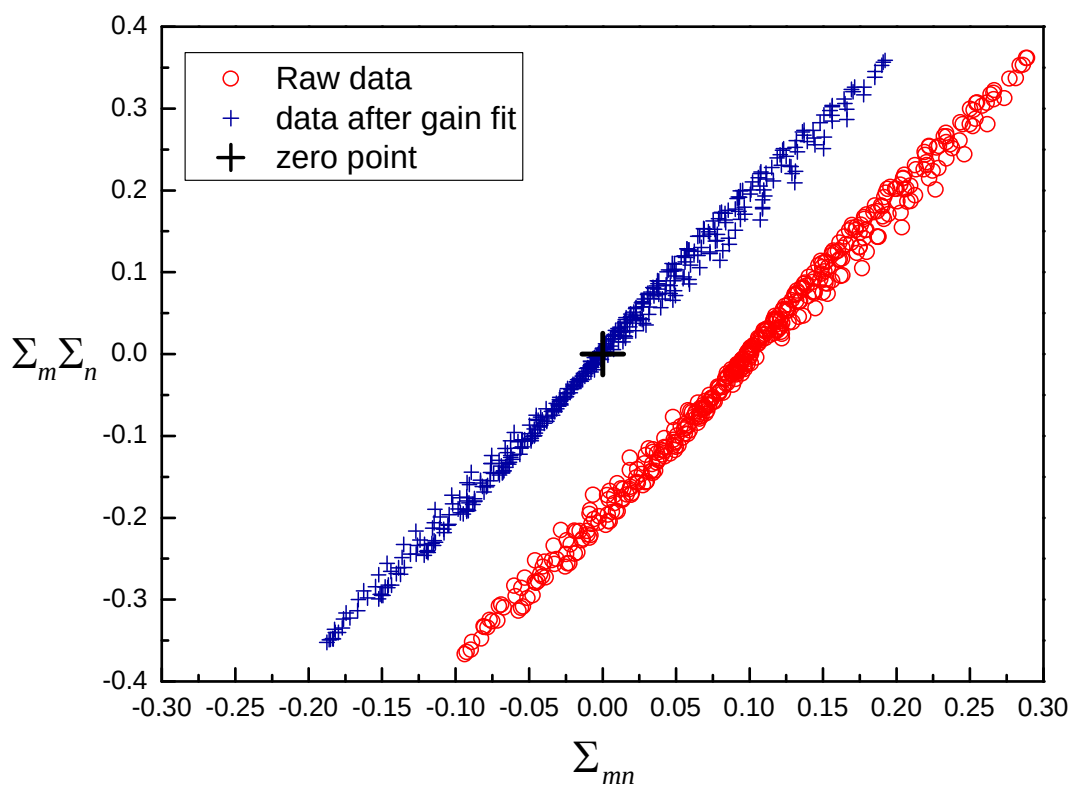

Fig. 4. $\Sigma_m\Sigma_n$ vs $\Sigma_{mn}$ for a calibration data at LA-BD-BPM03.

To verify the effectiveness above method, the table 1 shows the main geometric calibration parameters change of the LA-BD-BPM03 before and after gain fitting. Compared to the geometric coefficient before gain fitting, the geometric coefficient are closer to the theoretical value 7.55mm after gain fitting. Thus, the above method is effective.

Table 1. The change of calibration parameters before and after gain fitting

| | *before gain fitting* | | *after gain fitting* | |
|---|---|---|---|---|
| Position | x | y | x | y |
| Offset/mm | -0.19 | -0.15 | -0.13 | -0.01 |
| Geometric coefficient /mm | 7.60 | 7.41 | 7.60 | 7.45 |

Gains for 19 BPMs are shown in Fig. 5. The distribution of fitted gains is shown in Fig. 6. We can see most electrodes gain errors are between 0.9 and 1.1. Note that the average value of parameter $\tilde{k}_{mn}$ is 0.530, which is a little bit larger than the theoretical value.

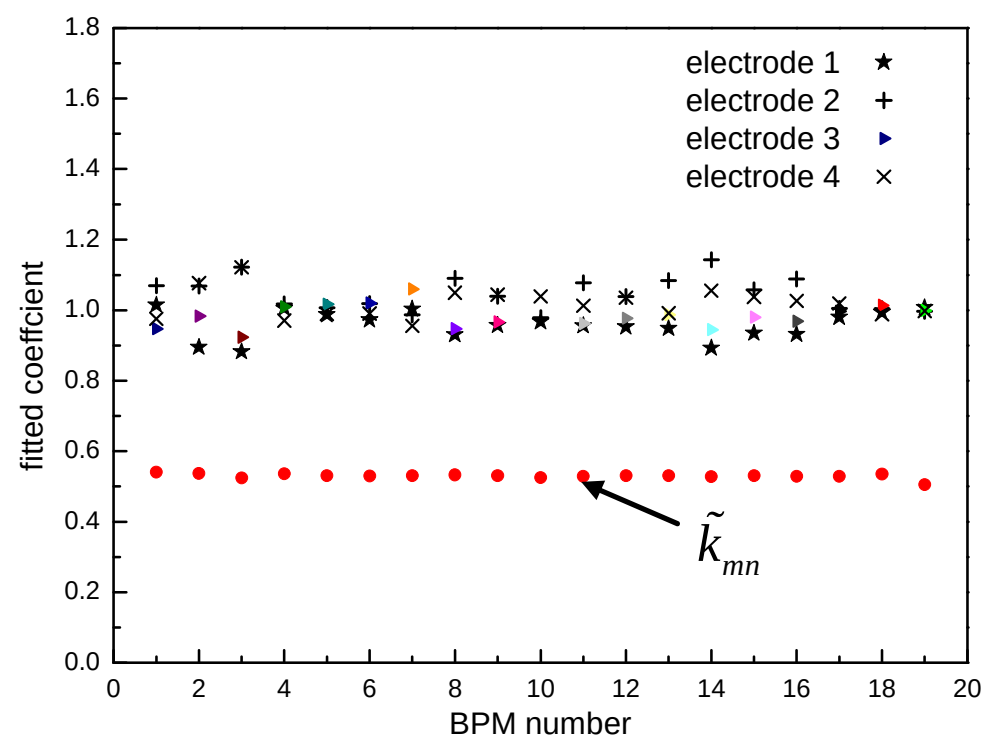

Fig. 5. Fitted gains and parameter $\tilde{k}_{mn}$ from calibration data for all 19 injector beam position monitors.

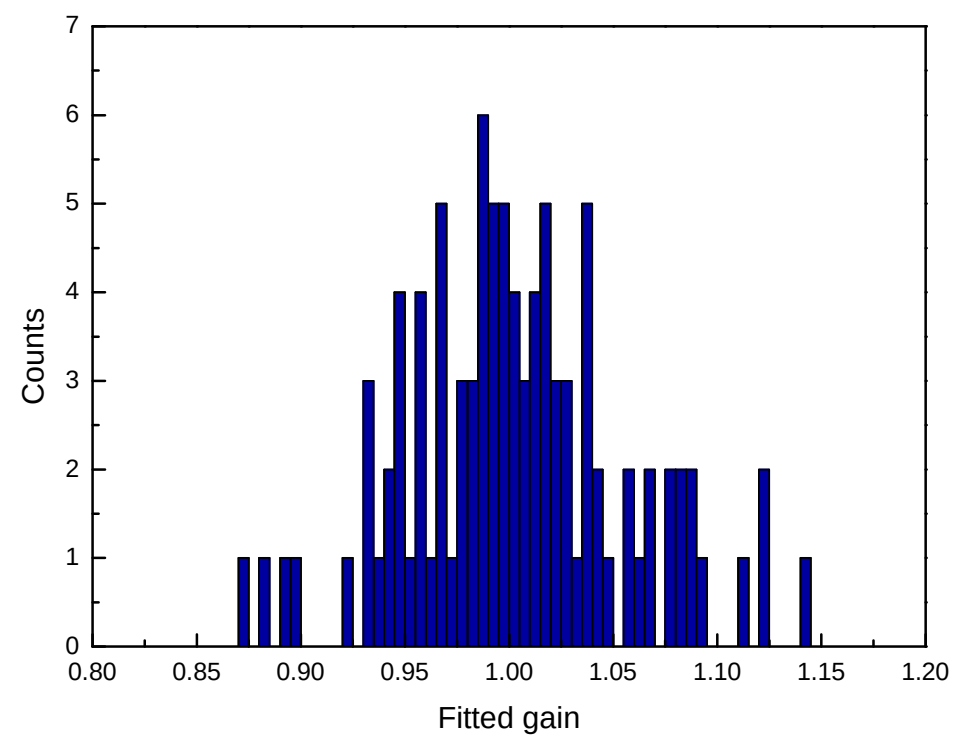

Fig. 6. Distribution of fitted gains for the data plotted in Fig. 5.

## 7 Conclusion

We have derived a relationship among the intensities of the four electrodes of orthogonal symmetrical type beam position monitor. The relationship is better than the previous study because it is irrelevant to the beam charge and the related coefficient can be theoretical calculated. We analyze the effect of electrodes coupling on the relationship. We show how the relationship can be used to make a beam based measurement of the relative gains of the four electrodes. We have used the calibration data to fit the gain for all 19 injector beam position monitors. The standard deviation of the distribution of measured gains is about 5%, consistent with the specifications of the system electronics. We will use the real beam data of HLS II injector to fit the electrodes gain, this can be implemented as a part of the standard measurements of the HLS II injector BPM system.

# References

1. ZOU Jun-Ying, YANG Yong-Liang, SUN Bao-Gen, et al. Calibration of Beam Position Monitors in the Injector of HLS II. Proceedings of IPAC2013, Shanghai, China, 2013, 568-570
2. Rubin D L, Billing M, Meller R, et al. Beam Based Measurement of Beam Position Monitor Electrode Gains. Phys. Rev. ST Accel. Beams, 2010, **13**(9): 092802-1~092802-6
3. LI Peng, SUN Bao-gen, LUO Qing, et al. New Methods of Beam Position Monitors for Measurement of Quadrupole Component. High Power Laser and Particle Beams, 2008, **20**(4): 573-578
4. Olmos A, Pérez F, Rehm G. Matlab Code for BPM Button Geometry Computation. Proceedings of DIPAC 2007, Venice, Itally, 2007, 186-188
5. ZOU Jun-Ying, FANG Jia, SUN Bao-Gen, et al. Application of Libera Brilliance Single Pass at NSRL Linac BPM System. Proceedings of IPAC2011, San Sebastián, Spain, 2013, 1284-1286